\documentclass[12pt]{article}
\usepackage{amsmath}
\usepackage{amssymb}

\newcommand{\be}{\begin{equation}}
\newcommand{\ee}{\end{equation}}
\newcommand{\ba}{\begin{eqnarray}}
\newcommand{\ea}{\end{eqnarray}}

\newcommand{\x}{\mathbf{x}}

\title{D-dimensional Conformal Field Theories with anomalous dimensions as Dual Resonance Models \thanks{Dedicated to Professor Ivan Todorov on the occasion of his 75$^{\text{th}}$ anniversary}
  }
\author{G. Mack\\
II. Institut f\"ur Theoretische Physik, Universit\"at Hamburg
}

\date{\today}
\begin{document}

\maketitle 
\begin{abstract} 
An exact correspondence is pointed out between conformal field theories in $D$ dimensions and dual resonance models in $D^\prime$ dimensions, where $D^\prime$ may differ from $D$.  Dual resonance models, pioneered by Veneziano, were forerunners of string theory. The analogs of scattering amplitudes in dual resonance models are  called Mellin amplitudes; they depend on complex variables  $s_{ij}$ which substitute for the Mandelstam variables on which scattering amplitudes depend. The Mellin amplitudes  satisfy exact duality - i.e. meromorphy in $s_{ij}$ with simple poles in single variables, and crossing symmetry - and an appropriate form of  factorization which is implied by operator product expansions (OPE). Duality is a $D$-independent property.  The position of the leading poles in $s_{12}$
are given by the dimensions of fields in the OPE, but there are also satellites and the precise correspondence between fields in the OPE and the residues of these poles depends on $D$. Dimensional reduction and dimensional induction
$D\mapsto D\mp 1$ are discussed. Dimensional reduction leads to the appearence of Anti de Sitter space
\end{abstract}

 \setlength{\unitlength}{0.3mm}
 \section{Introduction} 
 In this talk, I report an exact correspondence between correlation functions of $D$-dimensional conformal field theories (CFT) with anomalous field dimensions and scattering amplitudes of dual resonance models \cite{frampton:duality:reprint} in $D^\prime$ dimensions, where $D^\prime$ can be chosen and need not equal $D$. The dual resonance models satisfy exact duality - i.e. meromorphy of the scattering amplitudes plus crossing symmetry - and factorization.  Duality is a $D$-independent property. The appropriate form of factorization follows from operator product expansions (OPE) \cite{wilson:OPE} and depends on $D$. In general, the dual resonance models have  satellites, i.e. to one field $\phi^k$ in the OPE there corresponds an integer spaced  sequence of poles (corresponding to resonances in the dual resonance models of old) labelled by $n =0,1,2,...$. The leading pole (n=0) determines the satellites ($n=1,2,3,...$) in a $D$-dependent way. When all the 4-point functions, including those for fields of arbitrary Lorentz spin,  satisfy duality and factorization, and if all 2-point functions are positive, the physical principle of a local conformal  quantum field theory (Wightman axioms \cite{PCT} or Osterwalder Schrader axioms for the Euklidean Green functions \cite{OS:I,OS:II}) are obeyed. In this talk I concentrate on scalar correlation functions.  An expanded treatment with full proofs is published in \cite{mack:cftI}. 
 
 Maldacena's correspondence between String theory on 5-dimensional Anti-de Sitter space and $N=4$ Super Yang Mills theory in 4 dimensions suggests the conjecture that all conformal field theories are string theories. This has not been proven. But the correspondence with dual resonance models is a result which goes in the same direction. Dual resonance models, pioneered by Veneziano \cite{Veneziano:model} were forerunners of string theory. They  share many of its features, including a pivotal role of 2-dimensional conformal symmetry as reparametrization invariance in a space of auxiliary variables. 
 In string theory it became reparametrization invariance of the world sheet.
 
 Work in the seventies by Ivan Todorov and his collaborators, including myself,
 \cite{dobrevEtAl:OPE,mack:dobrevClebsch,mack:dobrevBook,mack:groupth:1,mack:groupth:2} showed by group theoretical means how one can ensure validity of all the physical principles of conformal field theory except locality.
Locality took the form of a duality relation \cite{mack:duality}.  In the new approach, one tries to start from it.  

Given that there exist efficient methods to construct 2-dimensional conformal field theories, 
it is tempting  to ask: How much of the structure of $D$-dimensional conformal field theory is $D$-dependent? A related theme is dimensional reduction and dimensional induction.

\section{Question: What part of the structure of conformal field theories is $D$-independent?}
As our starting point we observe that for any $D\geq 2$, $n$-point correlation functions are given by functions of
 {\em the same}
number $\frac 1 2 n(n-3) $ of independent anharmonic ratios $\omega_i$. 

Let $\x_{ij}=\x_i-\x_j$, $i<j=1...n$. 
 
Special case $D>2, n=4$ :  $\omega_1\omega_2\omega_3 =1, $ 
$$
\omega_1=\frac {\x_{12}^2 \x_{34}^2}{\x_{13}^2 \x_{24}^2}, 
\quad \omega_2 = \frac {\x_{14}^2 \x_{23}^2}{\x_{12}^2 \x_{34}^2} ,
\quad \omega_3 = \frac {\x_{13}^2 \x_{24}^2}{\x_{14}^2 \x_{23}^2} 
$$
Consider Euklidean Green functions  of a CFT or correlation functions of commuting scalar Euklidean fields. 
The 4-point functions have the form 
\begin{eqnarray}
 G_{i_4,... ,i_1}(\x_4,...,\x_1)&=&<\varphi^{i_4}(\x_4), ...,  \varphi^{i_1}(\x_1)>  \nonumber \\
    &=&  \prod_{i>j}(\x_{ij}^2)^{-\delta_{ij}^0}F_{i_4...i_1}(\omega_1,\omega_2,\omega_3) \nonumber 
\end{eqnarray}
$\delta_{ij}^0=\delta_{ji}^0 $ depend on the dimensions $d_j$ of the fields $\varphi^{i_j}$:
 $\sum_j\delta_{ij}^0=d_i $.% \textcolor{blue}{Euklidean$\leftrightarrow$Minkowski} $\x_{ij}^2 = -x_{ij}^2$

\section{Locality or crossing symmetry}
In the Euklidean domain, locality of bosonic fields turns into 
 commutativity of Euklidean fields $\varphi^i(x)$ and implies symmetry of Euklidean correlation functions
$$ G_{i_4,...,i_1}(\x_4,...,\x_1) = G_{i_{\pi 4},...,i_{\pi 1}}(\x_{\pi 4},...,\x_{\pi 1})   $$
for all permutations $\pi$ of $1...n=4$. Permutations $\pi : i \leftrightarrow j$ act on harmonic ratios via
$\omega\mapsto \pi \omega$,  as follows
\begin{eqnarray}
&&(ij)=(12)\mbox{ or }(34): \  \omega_1 \mapsto \omega_2^{-1}, \   \omega_2\mapsto \omega_1^{-1},  \omega_3\mapsto \omega_3^{-1} \nonumber \\
&&(ij)=(13)\mbox{ or }(24): \  \omega_1 \mapsto \omega_3^{-1}, \   \omega_2\mapsto \omega_2^{-1},  \omega_3\mapsto \omega_1^{-1} \nonumber \\
&&(ij)=(14)\mbox{ or }(23): \  \omega_1 \mapsto \omega_1^{-1}, \   \omega_2\mapsto \omega_3^{-1},  \omega_3\mapsto \omega_2^{-1} \  \nonumber
\end{eqnarray}
Therefore, locality is equivalent to symmetry properties of $F_{i_4,...,i_1}(\omega )$, viz.
 $$ F_{i_4,...,i_1}(\omega )=  F_{i_{\pi 4},...,i_{\pi 1}}(\pi \omega ) .$$
\section{Operator product expansions}
In CFT in {\bf Minkowski space} (or on its $\infty$-sheeted covering $\mathcal{M}_D\simeq S^{D-1}\times \mathbf{R}$), Wilsons OPE can be partially summed \cite{ferraraEtAl:conformalAlgebraOPE} and the partially summed OPE converge on the vacuum $\Omega$ \cite{mack:OPE}, 
\begin{eqnarray}
&&\phi^i(-\frac 1 2 x)\phi^j(\frac 1 2 x)\Omega = \nonumber \\
&& = \sum_{k}\sum_a g^{ij}_{k,a} \tilde{Q}^a(\chi_k,-i\nabla_z; \chi_j, -\frac 1 2 x , \chi_i,\frac 1 2 x)\phi^k(z)\Omega|_{z=0} \nonumber 
\end{eqnarray}
with {\em kinematically determined} coefficients  $\tilde{Q}^a$ and coupling constants 
$g^{ij}_{k,a}$. $\chi_k=[l_k,d_k] $ indicate Lorentz spin $l_k$ and dimension $d_k$ of field $\phi^k$. 
$k$-summation is over all nonderivative fields.

{\em The CFT is completely determined by knowledge of coupling constants $g^{ij}_{k,a}$ and spin and dimension $l_k,d_k$ of all nonderivative fields $\phi^k$}. Consistency requires locality of all 4-point functions, including those of fields with arbitrary spin \cite{mack:duality}.  OPE imply positivity (unitarity) if all fields $\phi^k$ have positive 2-point functions. This in turn requires that all the fields $\phi^k$ have Lorentz spin and dimension which determine 
unitary positive energy representations of the conformal group \cite{mack:irreps} , and there are no ghosts. 

The main theme of this talk will be to extract  {\em some $D$-independent structural properties of CFT} from OPE, and clarify what remains $D$-dependent, by use of    a

\section{Mellin representation of correlation functions}
Remember the Mellin representation of functions $f(x)$ of real variables $x>0$:
$ f(x)= (2\pi i)^{-1} \int_{-i\infty}^\infty ds \tilde{f}(s)x^{-s} $

Inserting the Mellin representation of $F_{i_4,...,i_1}(\omega_1,\omega_2,\omega_3)$ in independent 
(Euklidean) harmonic ratios, e.g. $\omega_1, \omega_2$, and extracting some normalization factors $\Gamma(...)$ by convention, we get for the 4-point function
$$G_{i_4,...,i_1}(\x_4,...,\x_1) = (2\pi i)^{-2} \int d^2\delta \ M_{i_4,...,i_1}(\{\delta_{ij}\} )\prod_{i>j} 
\Gamma(\delta_{ij})(\x^2_{ij})^{-\delta_{ij}}$$
Integration is over the 2-dimensional surface of imaginary $\delta_{ij}=\delta_{ji}$, $1\leq i <  j \leq 4$ subject to $\sum _j \delta_{ij}=d_i$.  
And similarly for scalar $n$-point functions $G_{i_n,...,i_1}(x_n,...,x_1)$, with $\frac 1 2 n(n-3) $ integrations, as indicated below. $M_{i_n,...,i_1}$ are called Mellin amplitudes of the CFT.

\section{$n$-point Wightman functions, time ordered Green functions and Euklidean Green functions from the same Mellin amplitude}

Let $m=\frac 1 2  n(n-3)$. {\bf In Euklidean space} 
$$G_{i_n,...,i_1}(\x_n,...,\x_1) = (2\pi i)^{-m} \int d^m\delta \ M_{i_n,...,i_1}(\{\delta_{ij}\} )\prod_{i>j} 
\Gamma(\delta_{ij})(\x^2_{ij})^{-\delta_{ij}}$$%
{\bf In  Minkowski space} the Wightman functions become
\begin{eqnarray}
<\Omega,
\phi^{i_n}(x_n)...\phi^{i_1}(x_1)\Omega>&=&
(2\pi i)^{-m} \int d^m\delta \ M_{i_n,...,i_1}
(\{\delta_{ij}\} ) \nonumber \\
&&  \prod_{i>j} \Gamma(\delta_{ij})(-x^2_{ij}+i\epsilon x_{ij}^0)^{-\delta_{ij}}
\nonumber
\end{eqnarray} 
and the time ordered Green functions read
\begin{eqnarray}
<\Omega, T\{ \phi^{i_n}(x_1)...\phi^{i_1}(x_1)\}\Omega>&=&
(2\pi i)^{-m} \int d^m\delta \ M_{i_n,...,i_1}(\{\delta_{ij}\} ) \nonumber \\
&& \prod_{i>j} \Gamma(\delta_{ij})(-x^2_{ij}+i\epsilon )^{-\delta_{ij}}\nonumber
\end{eqnarray}
\section{Duality properties of Mellin amplitudes}% $M_{i_n,...,i_1}(\{\delta_{ij}\})$}
 
{\em From localiy}: Mellin amplitudes are {\em symmetric} under permutations $\pi$ of $1...n$,
$$  M_{i_n,...,i_1}(\{ \delta_{ij} \} ) =  M_{i_{\pi n},...,i_{\pi 1}}(\{\delta_{\pi i  \pi  j} \} ) $$
{\em From OPE}: Mellin amplitudes 
$M_{i_n,...,i_1}(\{ \delta_{ij} \} )$
are {\em meromorphic functions} of the (independent) variables $\delta_{ij}$, with {\em simple poles} in single variables  (e.g. $\delta_{12}$), at positions which are determined by the twist $d_k-l_k$ of the fields $\phi^k$ in the OPE and whose {\em residues are polynomials} in the other independent variables .

More precise statements are made below,  and compared  to properties of dual resonance models. 

\section{Solution of the constraints $\sum_j\delta_{ij}=d_i$, and pole positions \label{8}}
Let $p_i $, $i=1,...,n$ be  conserved $D^\prime$ dimensional "momenta" satisfying $p_i^2 = d_i$,  and
$ \sum_i p_i =0 $

Then $\delta_{ij}= -p_i\cdot p_j $ satisfy the constraint $\sum_j\delta_{ij}=d_i$.  ($D^\prime $ need not equal $D$). 

Define Mandelstam variables:  $$s_{jl}=(p_j+p_l)^2 = d_j +d_l -2\delta_{jl} $$
If the OPE $$\phi^{i_j}(x_j)\phi^{i_l}(x_l) \Omega = ... + \phi^k\Omega + ... $$
include a field $\phi^k$ of Lorentz spin (rank) $l_k$ and dimension $d_k$, 
 then the Mellin amplitude has a 
"leading" pole  in $\delta_{jl}$ at position $s_{jl}=d_k-l_k$  and "satellite poles" at $s_{jl}=d_k-l_k + 2n$,
$n=1,2,3,...$ 

 The polynomial residues are of $l_k$-th order and are  proportional to $g^{jl}_k$. They depend on 
$l_k$, $n$, differences of external dimensions including $d_j-d_l$, {\em and on $D$. } 
\section{Dual resonance models}
 Consider for instance  scattering of 2 spinless particles into 2 spinless particles.   The same analytic scattering amplitude $A(s,t,u)$ defines scattering in all 3 channels: 
\begin{eqnarray}
12 \mapsto 34&& (\mbox{c.m. energy})^2 = s \geq \mbox{max } (m_1+m_2)^2, (m_3+m_4)^2\nonumber\\
13\mapsto 24&&(\mbox{c.m. energy})^2 = t \geq \mbox{max } (m_1+m_3)^2, (m_2+m_4)^2
\nonumber\\
14\mapsto 23&&(\mbox{c.m. energy})^2 = u \geq \mbox{max } (m_1+m_4)^2, (m_2+m_3)^2 
\nonumber
\end{eqnarray} 
\setlength{\unitlength}{.9mm}
\begin{figure}[h]
 \begin{picture}(50,50)(-8,0)
 \thicklines
 \put(25,25){\circle{20}}
 %\put(0,0) {\circle*{2}}
 \put(0,50){\vector(1,-1){18}}
 \put(-6 ,52){$p_4, i_4$}
 \put(52,52){$p_1,i_1$}
 \put(-6,-4){$p_3,i_3$}
 \put(52,-4){$p_2,i_2$}
 \put(0,0){\vector(1,1){18}}
 \put(50,0){\vector(-1,1){18}}
 \put(50,50){\vector(-1,-1){18}}
 {\small
 \put(65,30){$s_{ij}=(p_i+p_j)^2, \ i\neq j$}
 \put(65,20){$s_{12}=s, \ s_{13}=t, \ s_{14}=u $} 
 \put(65,40){particles of types $i_1,...,i_4$}
 \put(65,10){$s+t+u=m_1^2+m_2^2+m_3^2+m_4^2$}
 }

 \end{picture}
 \end{figure}

 Dual resonance models furnished {\em meromorphic} ("narrow resonance")  approximations to $A(s,t,u)$ 
  with simple poles in $s\equiv s_{12},t\equiv s_{13},u\equiv s_{14}$ with polynomial residues.
 \section{Duality}
   Consider  {\em resonances in the $s$-channel}, 
  of type $k$ with spin $l_k$ and mass $m_k$ which can couple to (decay into) particles $i_1+i_2$ with strength $g_k^{12}$ and to $i_3+i_4$ with strength $g_k^{34}$.  The dual resonance models of old are based on a narrow resonance approximation, such that  the scattering amplitude is the sum of the contributions  of resonances,
\begin{figure}[h]
\setlength{\unitlength}{0.55mm}
\begin{picture}(100, 30)(-1,-15)
\thicklines
\put (0,-5){\Large $A(s,t,u)=\sum_k \bar{g}_k^{34}  g_k^{12}\frac {P_{l_k}(\cos \theta_{12})}{s-m_k^2} = \sum_k$ }
\setlength{\unitlength}{.35mm}
%\put(90,0){
\put(140,0){
\put(150,-5){\circle*{5}}
\put(200,-5){\circle*{5}}
\put(150,-5){\line(1,0){50}}
\put(150,-5){\line(-1,-1){20}}
\put(150,-5){\line(-1,+1){20}}
\put(200,-5){\line(1,-1){20}}
\put(200,-5){\line(1,+1){20}}
\put(170,0){$k$}
\put(200,-5){\line(1,-1){20}}
\put(222,15){$1$}
\put(200,-5){\line(1,-1){20}}
\put(222,-28){$2$}
\put(124,15){\small{$4$}}
%\put(-26,10){$4$}
\put(124,-28){\small{$3$}}
%\put(-26,-28){$3$}
}
\end{picture}  
\end{figure}

 This is an equality of analytic functions, valid not only for $s\geq (m_1+m_2)^2$.\\
The scattering angle $\theta_{12}$  in the 12-channel is a \mbox{polynomial in $t$ or $u$}
 
Using  {\em resonances in the $u$-channel} we have instead 
\begin{figure}[h]
\setlength{\unitlength}{.55mm}
\begin{picture}(100, 30)(-1,-15)
\thicklines
\put (0,-5){\Large $A(s,t,u)=\sum_k \bar{g}_k^{23}  g_k^{14}\frac {P_{l_k}(\cos \theta_{14})}{u-m_k^2} = \sum_k$ } 
\setlength{\unitlength}{.35mm}
\put(140,0){
\put(175,-15){\circle*{5}}
\put(175,+5){\circle*{5}}
\put(175,-15){\line(0,1){20}}
\put(175,-15){\line(-3,-1){30}}
\put(175,-15){\line(+3,-1){30}}
\put(175,5){\line(-3,+1){30}}
\put(175,5){\line(+3,+1){30}}
\put(130,-28){$3$}
\put(210,-28){$2$}
\put(130,13){$4$}
\put(210,13){$1$}
}
\end{picture}
\end{figure}

 {\em Duality} says: Both amplitudes are the same, i.e. both sums are equal.

A similar equality holds for the $t$-channel.  The zonal spherical functions $P_l$ depend on $D$.

While dual resonance models were only approximations to scattering amplitudes, no approximation is involved in the meromorphy of Mellin amplitudes of conformal field theories. It is an exact property which follows from OPE. 
 
\section{Factorization properties}
 
  Consider first  the dependence of the amplitudes $A(s,t,u)=A_{i_4,...,i_1}(\{ s_{ij}\}) $  for scattering $i_1+i_2\mapsto i_3+i_4 $ on particle types $i_4,...,i_1$. 

  Duality guarantees symmetry under permutations $\pi$ 
  $$ A_{i_4 ,... ,i_1}(\{ s_{ i  j}\}) =A_{i_{\pi 4},_ ,... ,\i_{pi_1}}(\{ s_{\pi i \pi j}\}).$$
  The contribution of a $s$-channel resonance is the product of a {\em factorizing expression 
  $\bar{g}^{34}_kg^{12}_k$ which carries the dependence on $i_4,...,i_1$}, times a kinematically determined factor.
  More generally, for $2\mapsto n-2$ particles,
  \setlength{\unitlength}{.37mm}
%  \begin{picture}(100,60)
\begin{figure}[h]
  \begin{picture}(100,60)
\put(50,0){
  \thicklines
     \put(45,30){\oval(40,20)} 
  \put(27,38){\line(-2,1){20}}
  \put(26,33){\line(-4,1){20}}
  \put(26,27){\line(-4,-1){20}}
  \put(27,22){\line(-2,-1){20}}
  \put(33,31){\circle*{1}}
  \put(33,29){\circle*{1}}
  \put(0,45){$n$}
  \put(65,35){\line(2,1){20}}
  \put(65,25){\line(2,-1){20}}
  \put(87,40){$1$}
  \put(87,10){$2$}
  %\put(0,10){$n-2$}
  \put(100,23){  {\LARGE $=   \sum_k $}}	
  \put(150,0){
      \put(45,30){\oval(40,20)} 
  \put(27,38){\line(-2,1){20}}
  \put(26,33){\line(-4,1){20}}
  \put(26,27){\line(-4,-1){20}}
  \put(27,22){\line(-2,-1){20}}
  \put(33,31){\circle*{1}}
  \put(33,29){\circle*{1}}
  \put(0,45){$n$}
%  \put(65,30){\circle*{5}}
  \put(65,30){\line(1,0){30}}
  \put(77,35){$k$}
  \put(105,30){\circle{20}}
  \put(114,35){\line(2,1){20}}
  \put(114,25){\line(2,-1){20}}
  \put(135,40){$1$}
  \put(135,10){$2$}
  }
}
  \end{picture}
  \end{figure}
 the contribution of a resonance $k$ {\em factorizes} into a $3$-point amplitude times a $(n-1)$-point  amplitude.  And similarly for $m\mapsto n $ particles. 
\section{Comparison with properties of Mellin amplitudes $M_{i_4,...,i_1}(\{ s_{ij} \})$}
The poles of scattering amplitudes in dual resonance models as functions of the Mandelstam variables 
$s_{12}$ are determined by the masses squared $m_k^2$ of resonances in the $12$-channel. 

A corresponding statement holds for the Mellin amplitudes of CFT's. The Mellin amplitudes, considered as functions of the Mandelstam variable $s_{12}=d_1+d_2-2\delta_{12}$ has poles whose positions are determined by the twist $d_k-l_k$ of the fields in the OPE (with dimension $d_k$ and Lorentz spin $l_k$).
Summing up, there is a {\em correspondence}
$$s_{ij} = d_i+d_j -2\delta_{ij}, \qquad m_k^2=d_k-l_k.$$
\begin{itemize}
\item The meromorphy properties are the same. There are simple poles in individual variables  $s_{ij}$
\item The positions of the poles are the same, $s_{ij}=m_k^2$ (indep. of $m_i,m_j$) if there is a field $\phi^k $ with spin $l_k$ and dimension $d_k$ in the OPE of $\phi^i\phi^j$. In addition there are satellite poles at $s_{ij}=m_k^2+2n$, $n=1,2,...$.
\item The poles come with polynomial residues $P_{l_k}$ of degree $l_k$ which are related to zonal spherical functions. They depend on $D, n$ and differences of dimensions like $d_i-d_j$  and are not identically the same as in dual resonance models.
\item The residues of the leading poles factorize. The residues of the satellite poles are determined by the residues of the leading poles.
\end{itemize}

\subsection{Remark} $m_k$ is the lowest possible  {\em energy} of particle $k$. $d_k$ is the lowest possible  {\em conformal energy} (Eigenvalue  of conformal Hamiltonian $H$) in irreps.  $\mathcal{H}^{[l_k,d_k]}$ spanned by 
$\phi^k\Omega $, for scalar fields $\phi^k$.
\section{The analog of Regge trajectories}
 In {\em dual resonance models}, the  particles lie on Regge trajectories $K$, with masses
$$ m_k^2=\alpha^K(l_k) $$
In the simplest  models the trajectories are linear
$$ \alpha^K = \alpha_0 + \alpha^\prime l_k , \qquad \mbox{ $l_k$ increases  in steps of $2$.} $$
In {\em soluble models of CFT} there are trajectories
$$ d_k= \alpha_0 + l_k + \sigma_k $$
$\sigma_k=\mbox{anomalous part of the dimension}$. Hence
$$ m_k^2 \equiv d_k-l_k = \alpha_0 + \sigma_k $$
If the anomalous part $\sigma_k$ of the dimensions were $0$, poles would fall on top of each other.
But in the models, $\sigma_k$ increases  with $l_k$ to  limit $2\Delta$ or to $\infty$. Thus we get
{\em rising trajectories which are  approximately linear, with slope $0$} 

\section{CFT models with an expansion parameter}

In CFT, fields of dimension $d<\frac D 2 $ are called {\em fundamental fields}. The existence of  fundamental fields does not destroy  duality. But it entails a special feature:  The absence of Ferrara-Gatto-Grillo shadow poles \cite{ferraraEtAl:conformalAlgebraOPE}: The presence of a  scalar field of dimension $d$ in OPE implies that 
there is no field in the OPE of dimension $D-d$.
 
\subsection{$\phi^3$-theory in $D=6+\epsilon$ dimensions} 
The operator product expansions in $\phi^3$-theory in $6+\epsilon$ dimensions were worked out to lowest order in $\epsilon $ long ago  \cite{mack:shortDistance} .
 Written in a schematic way, the OPE read 
$$\phi\phi\Omega= c_\phi \mathbf{1}\Omega + (\phi + \sum_{l=2,4,...}  \phi_{\mu_1...\mu_l})\Omega $$
There is a fundamental field $\phi$. It has dimension $d= \frac {D-2}{2}+\Delta$,
$$ \Delta= \frac 1 {18}\epsilon + ...$$

In addition there are traceless symmetric tensor fields of even rank $2,4,...$. These 
fields $ \phi_{\mu_1...\mu_l}$ have dimensions $d_l= D-2+l+\sigma_l$, 
$$\sigma_l= 2\Delta -  \frac{4}{3(s+2)(s+1)}\epsilon + .... $$
This formula specializes to $d_0=D-d$, and the absence of the corresponding field in the OPE is in agreement with the absence of shadow poles of Ferrara et al. . Let us interpret 
 $d_l=2d + l - \mbox{binding energy}$,  and $2d$ as the energy of constituents. Then $\mbox{binding energy} \mapsto 0 $ as $l\mapsto \infty$.
A general proof of this fact for scalar field theories was given by Callan and Gross \cite{Callan:1973pu}. 

It follows from this result that
the poles  of the Mellin amplitude at $s_{12}=D-2+\sigma_l$ have a {\em limit point} at $s_{12}=2d$.
\subsection{$\mathcal{N}=4$ SUSY Yang Mills theory in $4$ dimensions}
$$ \sigma_l \sim \gamma \ln l \mbox{ as } l\mapsto \infty  ,  \qquad \gamma=\mbox{cusp anomaly} $$
The cusp anomaly has recently been computed \cite{bassoEtAl:cusp}. 

The poles at $s_{12}=D-2+\sigma_l$ have no limit point as $l\mapsto \infty$\\
Interpretation: constituent fields have $\infty$ dimension. 

\section{Dimensional Reduction \& Appearence of Anti de Sitter space}

The conformal group $G=SO(D,2)$ of Minkowski space is not simply connected, because the maximal compact   subgroup $K=SO(D)\times SO(2)$ is not simply connected. Its universal covering is $Spin(D)\times \mathbf{R}$.

The Hilbert space $\mathcal{H}$ of a CFT carries a unitary representation of $\tilde{G}=$universal covering of $G$. Its center is infinite, isomorphic to  $\mathbf{Z}_2\times \mathbf{Z}$ 

Space time $\mathcal{M}_D$ must be a homogeneous space $\tilde{G}/H$.

In the costumary conformal field theories (and also in their generalizations \cite{mack:deRiese})
$H$ contains the connected component of the socalled maximal  parabolic subgroup $P$ of $\tilde {G}$, which in turn 
contains the rotation group $U\simeq Spin(D-1)\subset \tilde{K}$. There are then two main possibilities
\begin{enumerate}
\item $H=P$: $\quad \mathcal{M}_D= \tilde{K}/U \simeq S^{D-1}\times \mathbf{R}$\\
 $\mathcal{M}_D$ admits a $\tilde{G}$-invariant causal ordering (i.e. it is a hyperbolic space)\cite{mack:luescher:global}\\
 fields $\phi^k$ can have {\em anomalous dimensions}.
\item $ H=P\times \Gamma$:  $ \mathcal{M}_D$=compactified Minkowski space\\
$\mathcal{M}_D$ has closed timelike curves. \\ ($\Gamma$ is a discrete group containing $\mathbf{Z}$, it depends on whether $D$ is even or odd). Fields $\phi^k$ have half integral dimensions.
\end{enumerate}
In this talk I focus on the first possibility.
\subsection{Manifestly conformal covariant formalism} 

Following Dirac \cite{Dirac36}, coordinatize points $x$ on (the two fold cover of)  compactified Minkowski space  by rays of lightlike vectors
 $\xi=(\xi_0...\xi_{D-1},\xi_{D+1},\xi_{D+2}) $ in $D+2$ dimensions,  $\xi \sim \lambda \xi $, $\lambda >0$. 
$$ \xi_0^2 - \xi_1^2 - ... -\xi_{D-1}^2 - \xi_{D+1}^2 + \xi_{D+2}^2 = 0 . $$
$x_\mu=\xi_\mu/\kappa \qquad \kappa = \xi_{D+1}+\xi_{D+2}$ for $\mu=0...D-1$.
Elements of $SO(D,2) $ act on $\xi$ as pseudorotations. 
\subsection{Orbits after dimensional reduction $D\mapsto D-1$.}
 Restrict $G$ to $SO(D-1,2)$, 
and correspondingly for its covering $\tilde G$. 

$\mathcal{M}_D$ decomposes into orbits
$$ \mathcal{M}_D = AdS_D \cup \mathcal{M}_{D-1}\cup AdS_D $$
$AdS_D$=universal  cover of $D$-dimensional Anti de Sitter space.
{\em We see that $AdS_D$ and its universal covering are orbits - i.e.  homogeneous spaces  - of the  $D-1$-dimensional conformal group.}

 $\mathcal{M}_{D-1}$ is the common boundary of the two $AdS$ spaces: $$\mathcal{M}_D \supset \mathcal{M}_{D-1}=\{ x^{D-1}=0\}= \{\xi_{D-1}=0\}$$This is seen as follows.  $\xi_{D-1}$ is $SO(D-1,2)$-invariant. Distinguish $\xi_{D-1}<0, \  \xi_{D-1}=0,\  \xi_{D-1}> 0$. Scale to $\xi_{D-1}=-1, \ \xi_{D-1}=0,\  \xi_{D-1}=1$ respectively. If $\xi_{D-1}\neq 0 $ then
 $$ \xi_0^2 - \xi_1^2 - ... -\xi_{D-2}^2 - \xi_{D+1}^2 + \xi_{D+2}^2 = 1 . $$
 after scaling. This is Anti de Sitter space.
 
 On compactified Minkowski space, $\xi $ and $-\xi$ are identified, therefore the two $AdS$-spaces are   identified. 

\subsection{Dimensional reduction of a CFT}

In the manifestly covariant formalism, Wightman functions (WF)
$$ W_{i_n,...,i_1}(\xi^n,...,\xi^1)=\kappa_n^{-d_n}...\kappa_1^{-d_1}<\Omega \phi^{i_n}(x_n)...\phi^{i_1}(x_1)\Omega> $$
are multivalued functions of $\xi_i$.  The Mellin representation becomes
$$ W_{i_n...i_1}(\xi^n,...,\xi^1) = (2\pi i)^{-m} \int d^m\delta \ M_{i_n... i_1}(\{\delta_{ij}\})
\prod_{i>j} \Gamma(\delta_{ij})(2\xi^i\cdot \xi^j)^{-\delta_{ij}}$$
$m=\frac 1 2 n(n-3)$. The restriction to $\xi_{D-1}=0$ exists, is invariant under the restricted conformal group, and is given by identically the same formula with the understanding that $\xi^i_{D-1}=0$. Hence
{\em  the dimensionally reduced CFT has the same Mellin amplitude $M$}.

\section{Dimensional induction of CFT's}
A tempting idea could be as follows.  Construct the dimensional reduction of a $3$-dimensional CFT as a 2-dimensional CFT, compute its Mellin amplitude, and use it to write down the correlation functions 
of the 3-dimensional theory. And similarly for higher dimensions. The question arises: What special properties are required of the lower dimensional theory? The answer is that it must have appropriate multiplets of fields. 

\subsection{Multiplets of fields} 
Conformal OPE involve nonderivative fields. 
But not all derivatives of fields in $D$ dimensions  (evaluated at $x_{D-1}=0$)  are derivatives in $D-1$ dimensions. Ordinary derivatives 
$\partial_{D-1}...\partial_{D-1}\phi^{k}(x)|_{x_{D-1}=0}$
do not transform right, but
one can 
use the Bargmann-Todorov homogeneous differential operator $D_A$ on the cone $\xi^2=0$ which transform as vectors \cite{todorov:bargmann}. In this way one gets field multiplets 
$$ \phi^k_{,n}= D_{D-1}...D_{D-1}\phi^i(\xi )|_{\xi_{D-1}=0} $$ ($n=0,1,2,...$, $n$ factors $D_{D-1}$). 
The missing generators $J_{AB}$, $A=D-1$ of $SO(D,2)$ act as generators of an "internal" symmetry. It can map $\phi\mapsto D_{D-1}\phi$.
\subsection{Does the 2-dimenional theory have $\infty$ conformal symmetry?}
The tentative answer is yes. One cannot expect that the 2-dimensional stress energy tensor $T$ is among the reduced fields. But the Wightman functions with  $T's$ can be computed from the Wightman functions without $T$'s by exploiting the  known commutation relations of $T$ with fields $\phi^i$ in 2 dimensions. This is a standard  method: One splits T into positve and negative frequencies and shifts them to the left resp right  until they act on the vacuum $\Omega$ \cite{belavin:polyakov:1984}.

\section{Mellin amplitudes for CFT from string theory?}
Maldacena's correspondence between 4-dimensional Supersymmetric Yang Mills Theory and String Theory on 5-dimensional Anti-de Sitter space motivates the conjecture that all local $D\geq 2$ dimensional conformal field theories  are equivalent to string theories. If so, how are 
their correlation functions related? Here is an  educated guess which is supposed to recover the Mellin amplitude $M$ of the conformal field theory from expectation values $\langle ... \rangle $ in the string theory. It uses the momenta $p_i$ introduced in section \ref{8}. 
\begin{equation}
\delta(\sum p_i) M(\{-p_i\cdot p_j\}) = \langle \int dV\ e^{i\sum_i p_{i\mu}X^\mu(\sigma_i,\tau_i)}\rangle
\end{equation}
for bosonic string. 
$dV$ is a (2d- conformal) invariant volume element which integrates over$(n-3)$ of the arguments $(\sigma_i,\tau_i)$ on which the world sheet position $X^\mu$ depends. The others can be arbitrarily (distinct) prescribed. This is an old device. The prototype of such a formula can be found in Veneziano's 1970 Erice lectures \cite{veneziano:erice1970}. 

One may conjecture that physical requirements are fulfilled if the state spaces of the string theory and the CFT can be identified. In the Maldacena AdS/CFT correspondence, such an identification is known \cite{DHoker:2002aw}.

\section*{Acknowledgements} 

I would like to thank Volker Schomerus and Yuri Pis'mak for stimulating discussions. Ivan Todorov kindly informed me that Bakalov and Nikolov had independently studied dimensional induction.

\end{document}